\documentclass[5p,twocolumn,times,number]{elsarticle}

\usepackage{lineno,hyperref}
\usepackage{hyperref}
\modulolinenumbers[5]

\newcommand{\taon}{$\tau$-lepton }
\newcommand{\taons}{$\tau$-leptons }
\newcommand{\xmax}{\ensuremath{X_{\rm max}}}

\newcommand{\lsim}{\mathrel{\hbox{\rlap{\lower.75ex \hbox{$\sim$}} \kern-.3em \raise.4ex \hbox{$<$}}}}
\newcommand{\gsim}{\mathrel{\hbox{\rlap{\lower.75ex \hbox{$\sim$}} \kern-.3em \raise.4ex \hbox{$>$}}}}

\journal{Journal of \LaTeX\ Templates}

\begin{document}

\begin{frontmatter}

\title{Space-based Extensive Air Shower Optical Cherenkov and Fluorescence Measurements using SiPM Detectors in context of POEMMA}

\author[1,2,3]{John F. Krizmanic\corref{cor1}%
\fnref{myfootnote} }
\ead{john.f.krizmanic@nasa.gov}

\author{for the POEMMA collaboration$^\dagger$}

\address[1]{University of Maryland, Baltimore County,
Baltimore, Maryland 21250 USA}

\cortext[mycorrespondingauthor]{Corresponding author}

\address[2]{Center for Research and Exploration in Space Science \& Technology II}
\address[3]{NASA/Goddard Space Flight Center Greenbelt, Maryland 20771 USA}


\begin{abstract}
Developed as NASA Astrophysics Probe-class mission, the Probe Of Extreme Multi-Messenger Astrophysics (POEMMA) is designed to identify the sources of ultra-high energy cosmic rays (UHECRs) and to observe cosmic neutrinos.  POEMMA consists of two spacecraft flying in a loose formation at 525 km altitude, 28.5$^\circ$ inclination orbits. Each spacecraft hosts a Schmidt telescope with a large collecting area and wide Field-of-View (FoV). A novel focal plane is employed that is optimized to observe both the UV fluorescence signal from extensive air showers (EASs) and the optical Cherenkov signals from EASs. In UHECR stereo fluorescence mode, POEMMA will measure the spectrum, composition, and full-sky distribution of the UHECRs above 20 EeV with high statistics along with remarkable sensitivity to UHE neutrinos. The POEMMA spacecraft are designed to quickly re-orient to a Target-of-Opportunity (ToO) neutrino mode to observe transient astrophysical sources with unique sensitivity. In this mode, POEMMA will be able to detect cosmic tau neutrino events above 20 PeV by measuring the upward-moving EASs for \taon decays induced from tau neutrino interactions in the Earth. In this paper, POEMMA's science goals and instrument design are summarized with a focus on the SiPM implementation in POEMMA, along with a detailed discussion of the properties of the Cherenkov EAS signal in the context of wide wavelength sensitivity offered by SiPMs.
A comparison of the fluorescence response between SiPMs and the MAPMTs currently planned for use in POEMMA will also be discussed, assessing the potential for SiPMs to perform EAS fluorescence measurements. 
\end{abstract}


\end{frontmatter}


\section{Introduction}

POEMMA was developed as one of the selected NASA Astrophysics Probes \cite{NASAprobeRA} concept studies to be provided to the 2020 Astronomy and Astrophysics Decadal Survey in support of the development of the Probe-class for future astrophysics missions. The details of the POEMMA science, instruments, spacecraft, and mission are detailed in the POEMMA NASA Probe Study report\cite{POEMMAnasaReport} as well as other publications \cite{2019ICRC...36..378O,2019EPJWC.21006008K}. The POEMMA instrument and mission design  is built upon previous space-based UHECR, UHE, and very-high energy (VHE) neutrino instrument and mission development, including the OWL study \cite{OWL}, JEM-EUSO \cite{JEM-EUSO}, EUSO-SPB1 \cite{EUSO-SPB1}, EUSO-SPB2 \cite{EUSO-SPB2} development, and the CHANT study \cite{CHANT}.

The POEMMA is designed to employ space-based observations of UHECRs and cosmic neutrinos to achieve significant increases in exposure for of these two messengers of extremely  energetic astrophysical phenomena. POEMMA's UHECR science goal is to provide a significant increase in UHECR exposure while also providing excellent angular, energy, and nuclear composition resolution for $E_{CR} \gsim 20 $ EeV and with full-sky coverage of the celestial sphere. In stereo UHECR mode, the POEMMA satellites are separated by 300 km and tilted slightly away from nadir for each to view a common area and stereoscopically measure the EASs from UHECRs. POEMMA's relatively fine pixel angular resolution (0.084$^\circ$ pixel FoV) coupled to the stereo geometric reconstruction provides high-quality measurements of the longitudinal development of EAS. POEMMA's UHECR measurement capability is summarized in Table~\ref{UHECRtab} based on detailed simulated performance reported in Ref.~\cite{PhysRevD.101.023012} demonstrating the remarkable POEMMA's UHECR measurement ability and the astrophysics reach enabled by the significant increase in the statistics of observed UHECRs at the highest energy scale over then entire sky. POEMMA's excellent UHE EAS stereo fluorescence measurement performance also allows for unparalleled sensitivity to UHE neutrinos as well as unique capabilities in fundamental physics measurements including the proton-proton cross-section at $\sqrt{s} \approx 320$ TeV and unique sensitivity to Super Heavy Dark Matter (SHDM) annihilation into photons and neutrinos.

\begin{table}
\centering
\caption{POEMMA UHECR  simulated measurement capabilities\cite{PhysRevD.101.023012}.}
\label{UHECRtab}       
\begin{tabular}{ll}
\hline
Parameter &  Performance   \\ \hline
 UHECR Stereo & 260,000 km$^2$ sr (50 EeV) \\
 Geometry Factor & 400,000 km$^2$ sr ($\ge 100$ EeV) \\ \hline
 Obs Duty Cycle & 13\%  \\ \hline
 Physics Energy Thres & 20 EeV \\ \hline
UHE Stereo Energy Res & $<19\%$ (50 EeV) \\ \hline
UHE Angular Res & $\le 1.2^\circ$ (50 EeV) \\ \hline
UHE \xmax Red & $\le 30$ g/cm$^2$ (50 EeV) \\ \hline
UHECR rejection factor & \\
for UHE neutrinos & $2 \times 10^{-4}$ \\ \hline
Sky Coverage & $\pm 20\%$ @ 50 EeV (1 Year) \\
Variability &  $\pm 10\%$ @ 50 EeV (5 Year) \\ \hline
 \end{tabular}
\end{table}

POEMMA is also optimized to measure the tau neutrino flux from transient astrophysical events by slewing the spacecraft to view the optical Cherenkov signals from upward EASs initiated by
\taons produced by $\nu_\tau$ interacting in the Earth with significant sensitivity for $E_\nu \gsim 20$ PeV. POEMMA 's ability to  quickly re-point ($90^\circ$ in $500$~s) each spacecraft to the direction of an astrophysical ToO source combined with the nature of the spacecraft orbits  provides an extraordinary capability to follow up transient alerts with the ability to survey of the full sky.  The POEMMA satellites also have the ability to reduce their separation to 25 km on a few hour time scale. This allows for the coincidence measurement within the same Cherenkov light pool for an upward-moving EAS from a \taon decay and effectively lowers the neutrino energy threshold. This flight maneuver is envisioned to be performed for good long-duration neutrino transient, such as a neutron star -- neutron star merger detected via gravity waves with a detected electromagnetic counterpart. The details of POEMMA's  simulated neutrinos sensitivity and astrophysics reach are provided in Refs.~\cite{2019PhRvD.100f3010R} and \cite{2019arXiv190607209V}, the latter demonstrating nearly an order of magnitude improved sensitivity for long-duration neutrino transients for $E_\nu \gsim 300$ PeV and over an order of magnitude improvement for short-duration neutrino transients for $E_\nu \gsim 100$ PeV compared to current ground based experiments, e.g. IceCube and the Pierre Auger Observatory.  This exceptional performance is based on the use of SiPMs for the measurement of the optical Cherenkov EAS signals over a large wavelength band.

\begin{figure}[h]
\centering
\includegraphics[width=0.99\columnwidth]{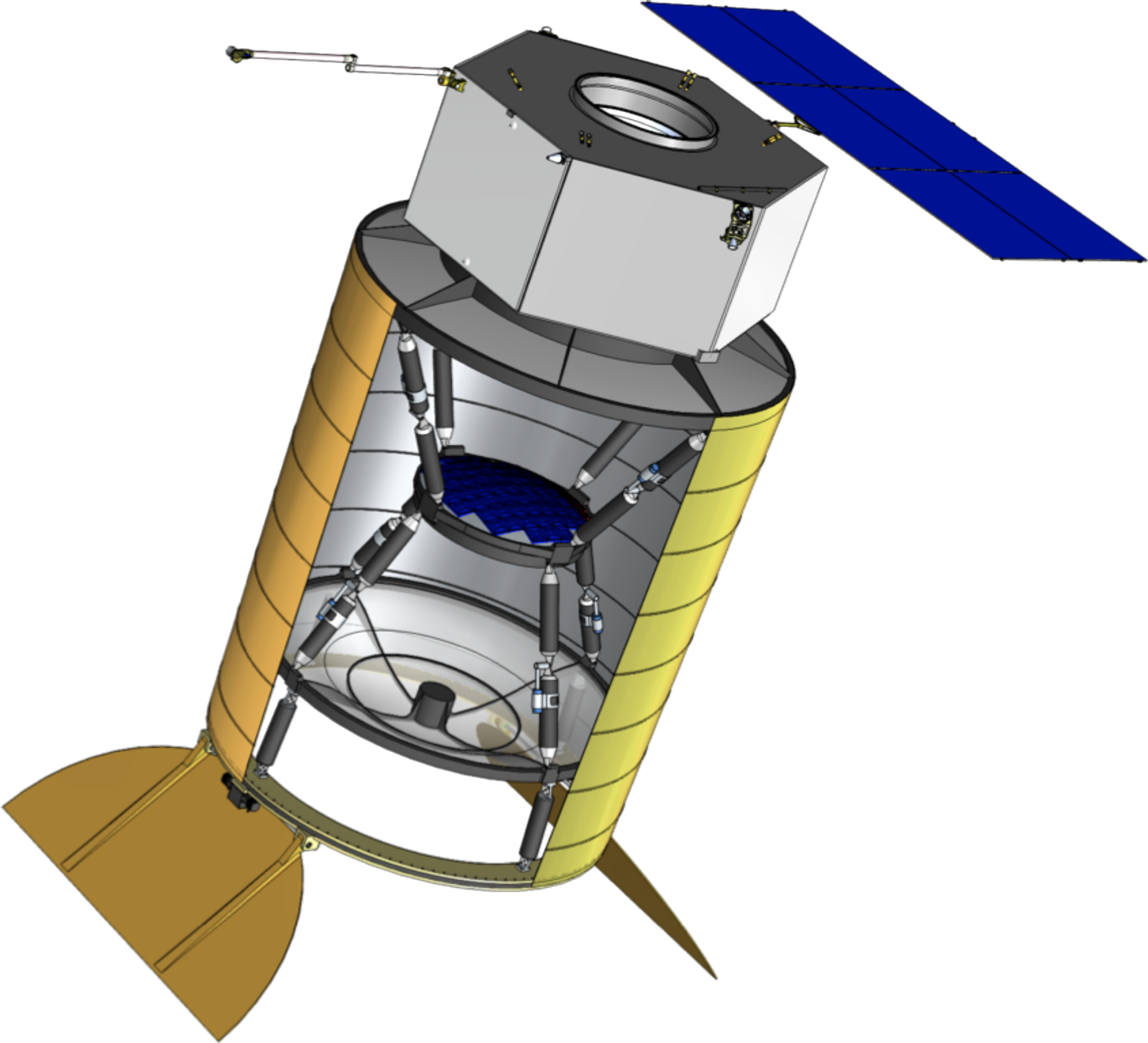}
\caption{A schematic of a POEMMA satellite looking downwards with the light doors open.  The light shield is cut away to view  the Schmidt telescope optics. Adapted from Fig. 17 in Ref. \cite{POEMMAnasaReport}.}
\label{POEMMAsatellite}       
\end{figure}

Each POEMMA instrument is comprised of a f/0.64 Schmidt telescope using a 3.3 meter diameter corrector lens, a 4-meter diameter monolithic primary mirror, and a 1.6 meter diameter focal surface. The details of the telescope and performance are in Tab.~\ref{OpticsTab}. A schematic of a deployed telescope looking downward with the light doors open is shown in Figure~\ref{POEMMAsatellite}. Each instrument also includes an IR camera, located slightly above the center of the corrector, to measure the cloud cover in the field of view  and calibration LEDs, located slightly underneath the outside edge of the corrector lens, used for continuous optics alignment verification.
Each POEMMA Schmidt telescope has a wide, full Field-of-View (FoV) of 45$^\circ$, and the effective area is nearly $6~{\rm m}^2$ on-axis decreasing to $\sim2~{\rm m}^2$ at the edge of the FoV. The point-spread-function (PSF) of the POEMMA optics have a RMS diameter 
no more than the 3~mm spatial linear pixel size of the photodetectors in the focal plane over the entire FoV. Thus the optical performance is well matched to the pixel FoV of $0.084^\circ$.  It is important to note that the optical performance is well matched for high-fidelity measurements of the longitudinal development of EASs, but is nearly a factor of 10,000 away from the diffraction limit in the near UV.  Thus the POEMMA Schmidt telescopes are more of a photometer given the imaging performance needed for UHECR and cosmic neutrino measurements.

\begin{table}
\centering
\caption{POEMMA Schmidt Telescope Specifications}
\label{OpticsTab}       
\begin{tabular}{lll}
\hline
Optics &  Schmidt &45$^\circ$ Full FoV   \\\hline
 & Primary Mirror & 4 meter diameter\\
 & Corrector Lens & 3.3 meter diameter \\
 & Focal Surface & 1.6 meter diameter \\ 
 & Pixel Size & $3 \times 3$ mm$^2$ \\ 
& Pixel FoV & 0.084$^\circ$ \\ 
 & Effective Area & $\sim$6 m$^2$ (on-axis) \\
 & & $\sim$2 m$^2$ (@ 22.5$^\circ$)  \\
 & PSF Diameter & $\lsim 3$ mm rms \\\hline
\multicolumn{3}{l}{Hybrid Focal Plane}  \\
PFC & MAPMT & 126,720 pixels \\
 &  Wavelength Band & 300 - 500 nm \\
 & Time Sampling & 1 $\mu$s \\
PCC & SiPM & 15,360 pixels\\ 
 &  Wavelength Band & 300 - 1000 nm \\
 & Time Sampling & 10 ns \\ \hline

\end{tabular}
\end{table}

\begin{figure}[h]
\centering
\includegraphics[width=0.99\columnwidth]{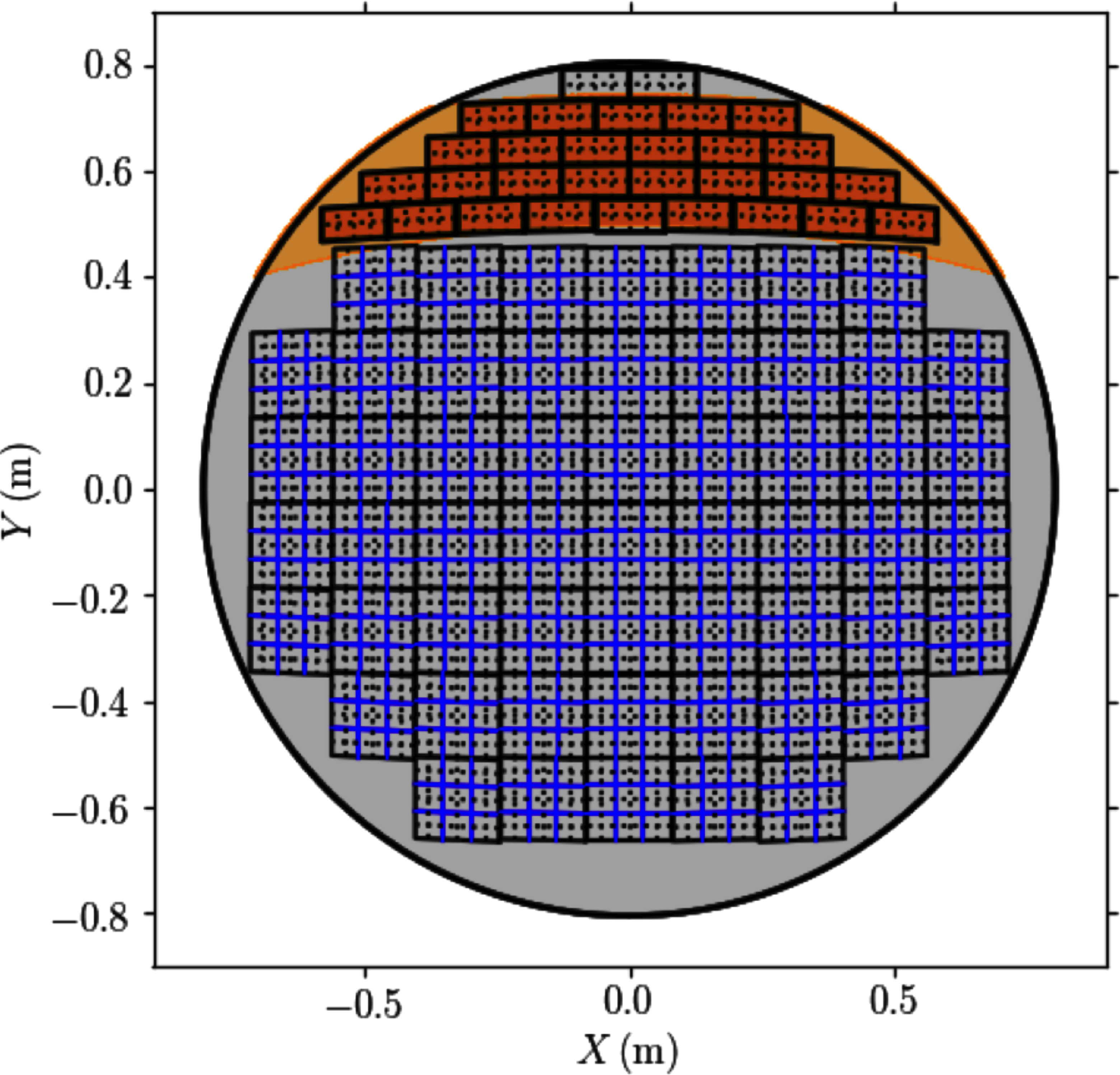}
\caption{The layout of a POEMMA Schmidt telescope focal surface with PDM and FSU modules. The 9$^\circ$ angular span of the the upper PCC segment is denoted by a 7$^\circ$ span (orange) with a  2$^\circ$ span above it. Adapted from Fig. 21 in Ref. \cite{POEMMAnasaReport} by Claire Guepin.}
\label{FocalPlane}       
\end{figure}

The focal surface in each POEMMA telescope is divided into two sections, a larger one, the POEMMA Fluorescence Camera (PFC), and a smaller one, the POEMMA Cherenkov Camera (PCC). The PFC is optimized for UHECR EAS longitudinal air fluorescence measurements while the PCC is optimized for the measurement of Cherenkov signals generated by upward-moving EASs. The PFC uses 55 Photo Detector Modules (PDMs) based on the JEM-EUSO instrument development \cite{2015ExA....40...19A} and each PDM consists of 36 64-channel MAPMTs. A BG3 filter is located on each MAPMT to constrain the wavelength to that of the UV fluorescence band (300 -- 500 nm) to minimize the effects of the atmospheric nightglow background (NGB) due to air glow. The PFC contains 26,720 $3\times3$ mm$^2$ pixels and will record signals using 1 $\mu$sec temporal sampling. 
The PCC consists of 30  focal surface units (FSUs) with each FSU consisting of a 512-channel array of silicon photomultipliers (SiPMs), whose broader wavelength response is better matched to inherent spectral variability of the  Cherenkov light measurement. 
In the PCC baseline design, silicon PIN pad detectors are located underneath each FSU SiPM array to reject the backgrounds in low-Earth orbit, including  cosmic rays and atmospheric `albedo' charged-particles. 
The FCC is sized to measure Cherenkov signals up to 9$^\circ$ away from the edge of the FoV of a telescope, as shown in Figure~\ref{FocalPlane}, based on an optimization for detecting the tau neutrino diffuse flux detailed in Ref.~\cite{2019PhRvD.100f3010R}.  Viewing the upward-moving EAS from 525 km altitude to 7$^\circ$ away from the Earth's limb translates to \taon Earth-emergence angle out to $\sim 20^\circ$ where the \taon emergence probability below 100 PeV is relatively high due to $\nu_\tau$ regeneration in the Earth Ref.~\cite{2019PhRvD.100f3010R}. Viewing 7$^\circ$ away from the Earth's limb also allows to survey the sky 2$^\circ$ above the limb to assess background events from the Cherenkov signal of UHECRS. The azimuth angular span of the PCC is $\sim 30^\circ$. The $9^\circ \times ~30^\circ$ viewing span of the FCC is more than sufficient for detecting ToO neutrino sources given the fact that the scale of angular error is defined by the Cherenkov angle, which is $\lsim 1.5^\circ$. The FCC contains 15,360 $3\times3$ mm$^2$ pixels, with custom electronics designed for recording signals using 10 nsec temporal sampling.  The front-end electronics for the PDMs and FSUs are located directly behind the focal surface with signals passed to the data acquisition electronics located underneath the primary mirror and in the satellite bus, which includes the power, communication, avionics, and  systems.

\section{Detection of the EAS Optical Cherenkov Light from \taon Decays}

Space-based measurement of the Cherenkov signal from upward-moving EASs created from neutrino interactions in the Earth offer a path to probe the  neutrino flux at lower neutrino energy scale due to the beamed nature of optical Cherenkov emission \cite{Krizmanic2011, CHANT} while also having a gargantuan neutrino target mass \cite{Domokos:1997ve,Domokos2}. The kinematics of tau neutrino charge-current interactions at and above the PeV scale \cite{Gandhi:1995tf,Block:2014kza} and the relatively long \taon decay lengths at high energy, $\gamma {\rm c} \tau$ for a \taon is nearly 50 km at 100 PeV, provides a unique signature for detecting these events \cite{Halzen:1998be,Beacom:2001xn,Bertou:2001vm,Feng:2001ue}, especially via EASs \cite{Fargion:2000iz, Bertou:2001vm, Bottai:2002nn,Fargion:2003kn}.
To first order, the flavor ratio of cosmic neutrinos, $\nu_e:\nu_\mu:\nu_\tau$, generated at a cosmological distance scale is expected to be close to $1:1:1$ due to neutrino oscillations \cite{Learned:1994wg}.

\begin{figure}[h]
\centering
\includegraphics[width=0.99\columnwidth]{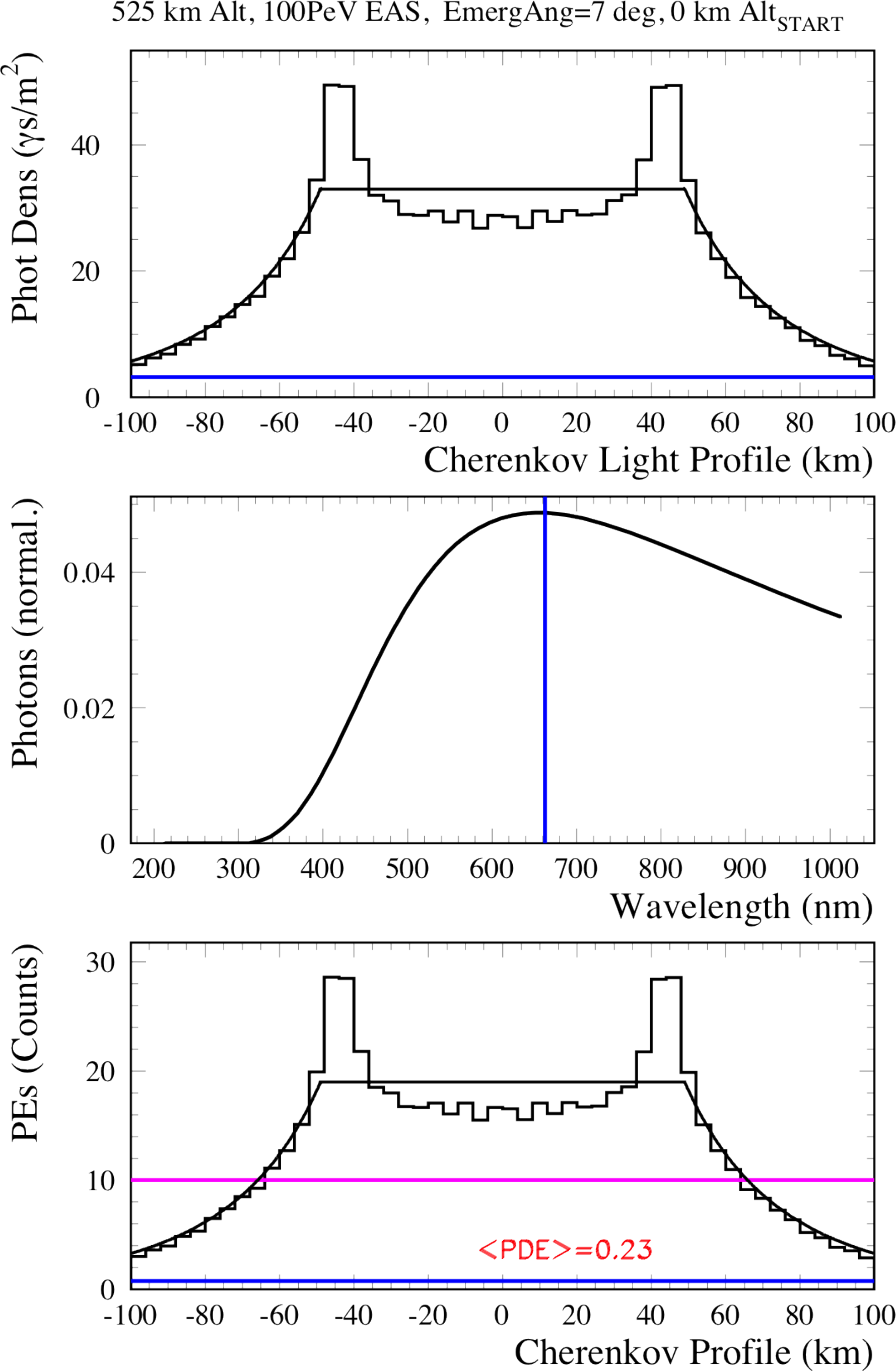}
\caption{The spatial profile of the optical Cherenkov signal (top), the Cherenkov spectrum at 525 km altitude (middle), and the photo-electron (PE) signal assuming the measured PDE for a Hamamatsu S14520-6050CN SiPM (bottom) for a 100 PeV EAS starting at sea level with $\beta_E=7^\circ$. The spectral weighted $<\rm{PDE}>$ obtained using the simulated Cherenkov spectrum is 23\%. The blue lines in the uppermost and lowermost 
plot show the average NGB signal in an $0.084^\circ$ FoV pixel assuming 10 ns sampling. A PE threshold of 10 is denoted by the magenta line.}
\label{cprofile}       
\end{figure}

\begin{figure}[h]
\centering
\includegraphics[width=0.99\columnwidth]{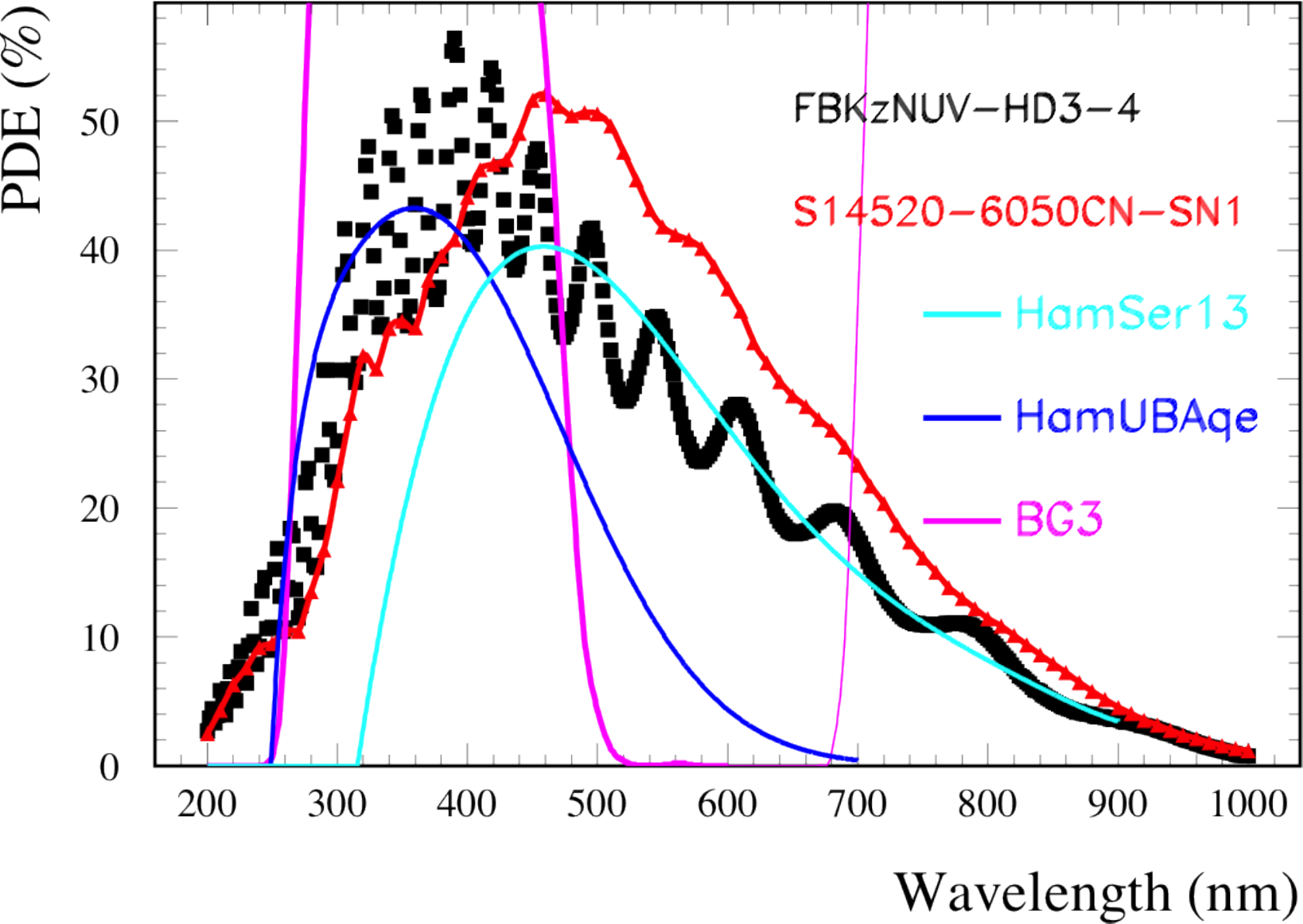}
\caption{Comparison of PDE response versus wavelength for various devices included the QE for a Hamamatsu ultra-bialkali PMT. The  bandpass of a Schott BG3 filter is also shown with the thicker magenta line highlighting the transmission for $\lambda \le 600$ nm and the thinner line for $\lambda > 600$ nm.}
\label{PDEcomp}       
\end{figure}

Key aspects of the optical Cherenkov signal viewed from low-Earth orbit are shown in Fig.~\ref{cprofile} for a simulated 100 PeV upward-moving EAS that started at sea level with an Earth-emergence angle of $\beta_E=7^\circ$.  In the uppermost plot, the Cherenkov photon density at 525 km  altitude is relatively flat between two horns, which are formed by the majority of particles around shower maximum, denoted as \xmax, when the Cherenkov angle about \xmax ~does not have much variation over the EAS development distance. Outside the horns, an exponential decrease is observed due to the angular spread of the electrons (and positrons) in the EAS, and these tails of the distribution fits well to an exponential parametric model\footnote{Thanks to Andrii Nerenov for the functional form.} and is shown overlaid on the plot. In the middle plot, the Cherenkov spectrum after atmospheric attenuation is presented and shows a peak near 660 nm. Below $\sim$310 nm, the Earth's ozone layer has negligible transmission and decimates the Cherenkov flux below this value.  Molecular (Rayleigh) and aerosol (Mie) scattering convolved with the $1/\lambda$ spectrum from the Cherenkov generation define the overall shape. In the lowermost plot, the Cherenkov intensity profile (top plot) is transformed into the PE signal profile assuming the parameters of POEMMA: e.g. 2.5 m$^2$ light collection area, a point-spread-function (PSF) of the optics that is matched to the angular spread of the Cherenkov signal (for POEMMA the pixel FoV is $0.084^\circ$), and the measured PDE curve of a Ham1420-6050CN SiPM \cite{OTTE2019283}, shown Fig.~\ref{PDEcomp}. Also shown in this figure is a comparison to the measured PDE response of an FBK\_NUV-HD3 SiPM \cite{OTTE2017106}
and the response of a Hamamatsu ultra-bialkali MAPMT and Series-13 SiPM as reported by the manufacturer \cite{HamUBA,Ham13}.
Note that the POEMMA neutrino sensitivity studies presented in Refs.~\cite{2019PhRvD.100f3010R} and \cite{2019arXiv190607209V} assumed a 20\% PDE for the Cherenkov signal and 10\% PDE for the NGB background based on the Hamamatsu Series-13 SiPM PDE response shown in Fig.~\ref{PDEcomp}.

\begin{figure}[h]
\centering
\includegraphics[width=0.99\columnwidth]{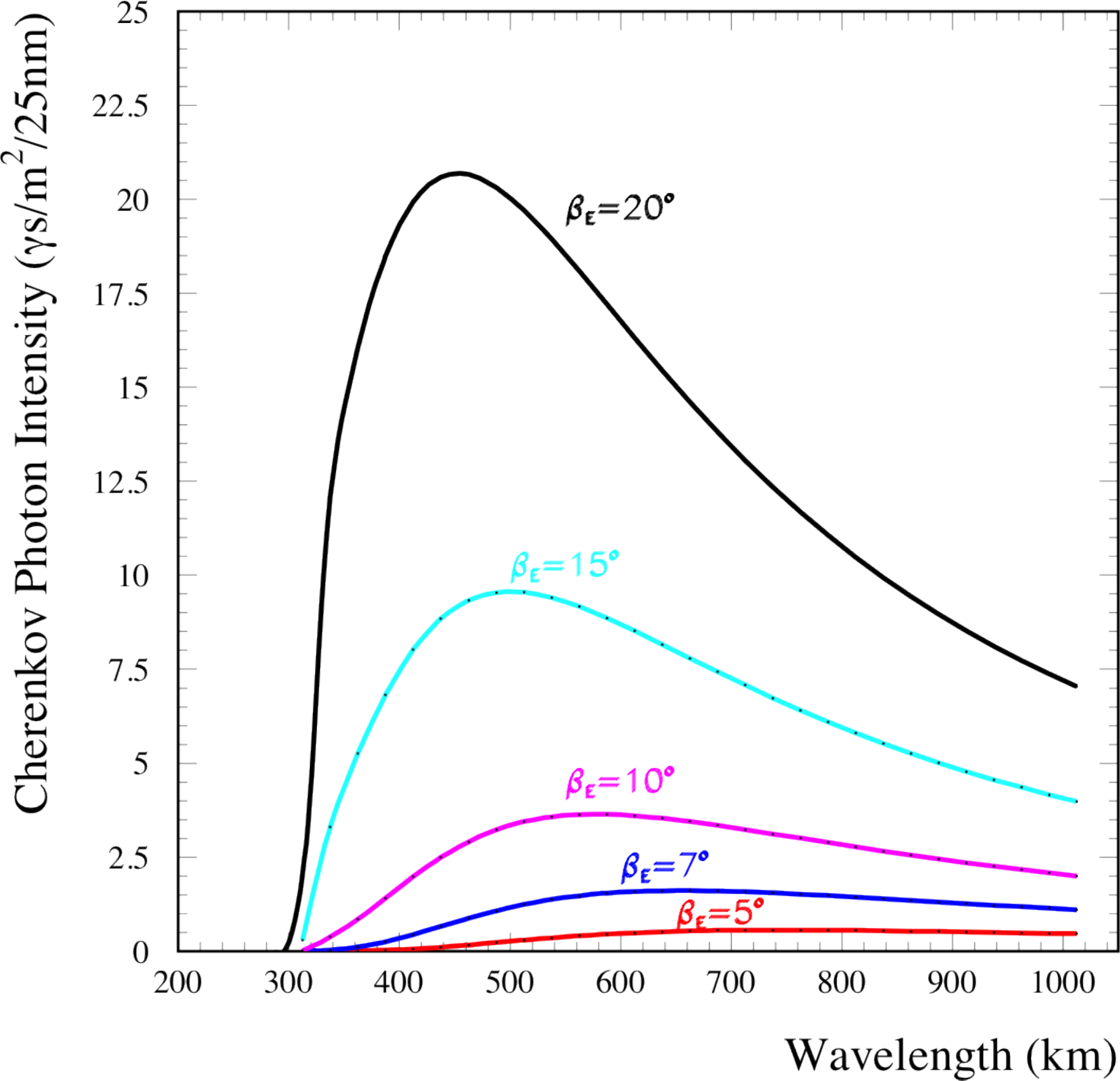}
\caption{The Cherenkov intensity spectrum, photons/m$^2$/25nm, for 100 PeV simulated EASs as a function of Earth-emergence angle ($\beta_E$.)}
\label{EASvsBeta}       
\end{figure}

Another key aspect of the EAS Cherenkov signal observed from space-based instruments is the large variability in the intensity and Cherenkov spectrum due to the atmospheric scattering and attenuation, especially due to aerosols. Fig.~\ref{EASvsBeta} presents the Cherenkov spectrum intensity (photons/m$^2$ per 25 nm wavelength band) for 100 PeV EASs viewed from 525 km starting their development at sea level as a function of Earth-emergence angle ($\beta_E$) of the \taon.  In the range  $1^\circ \le \beta_E \le 20^\circ$, the intensity varies by a factor of 20 while the peak of the Cherenkov spectrum moves to shorter wavelengths as $\beta_E$ increases.  The cause for both variabilities is due to aerosol attenuation in the atmosphere.  The attenuation is both a strong function of wavelength and altitude \cite{Elterman}, with the nominal atmospheric scale height $\sim$ 1 km. It should be noted that if conditions inject a significant amount of material into the stratosphere, such as from large volcanic eruptions (see Fig. 18.3 in Ref.~\cite{HGSE}) or from massive wild fires \cite{Yu587}, then the effects of the aerosol attenuation will extend to much higher altitudes for some time before returning the nominal aerosol profile.

\begin{figure}[t]
\centering
\includegraphics[width=0.99\columnwidth]{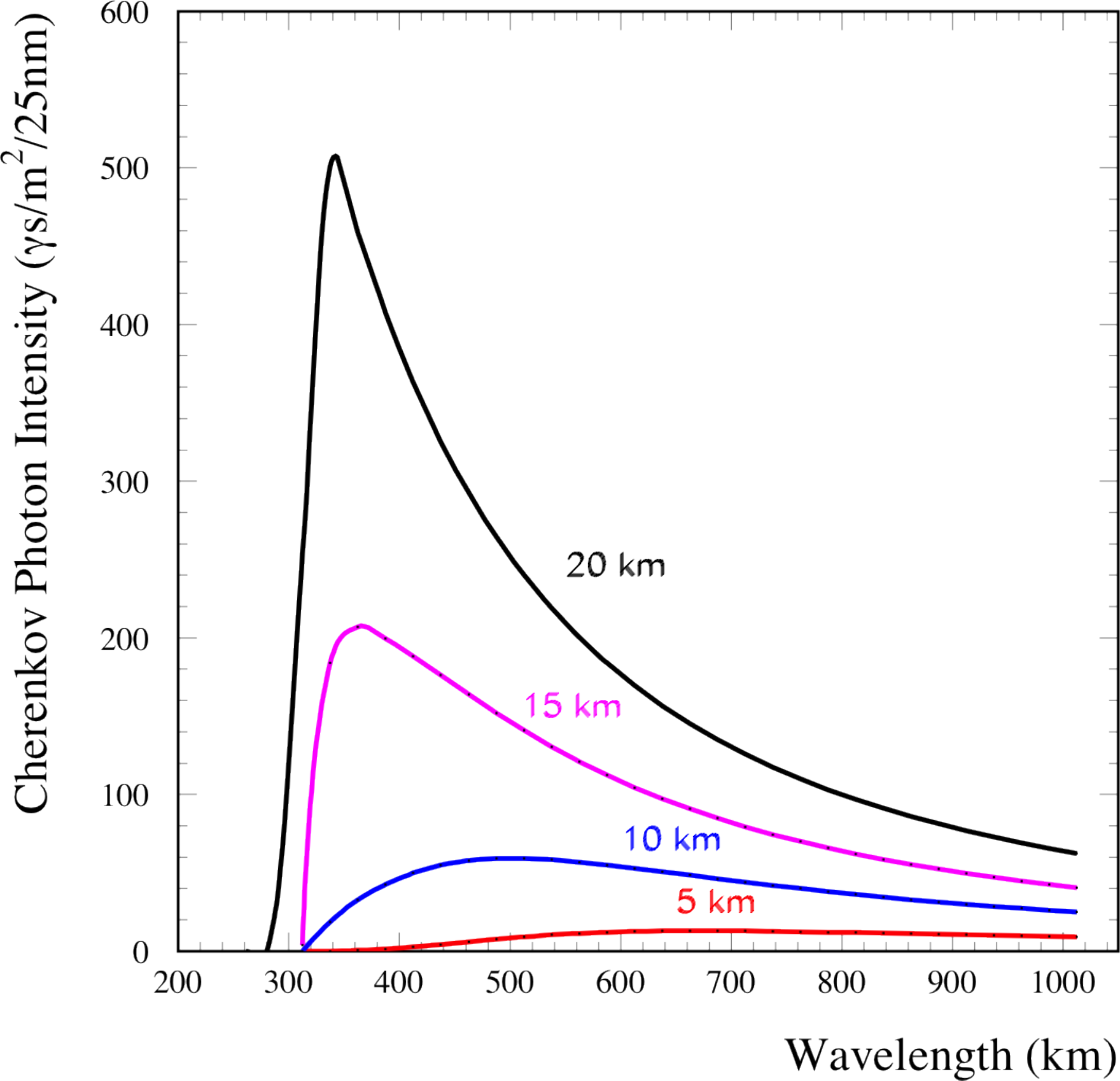}
\caption{The Cherenkov intensity spectrum, photons/m$^2$/25nm, for 100 PeV simulated EASs for $\beta_E=1^\circ$ as a function of EAS starting altitude.}
\label{EASvsAlt}       
\end{figure}

\begin{figure}[h]
\centering
\includegraphics[width=0.99\columnwidth]{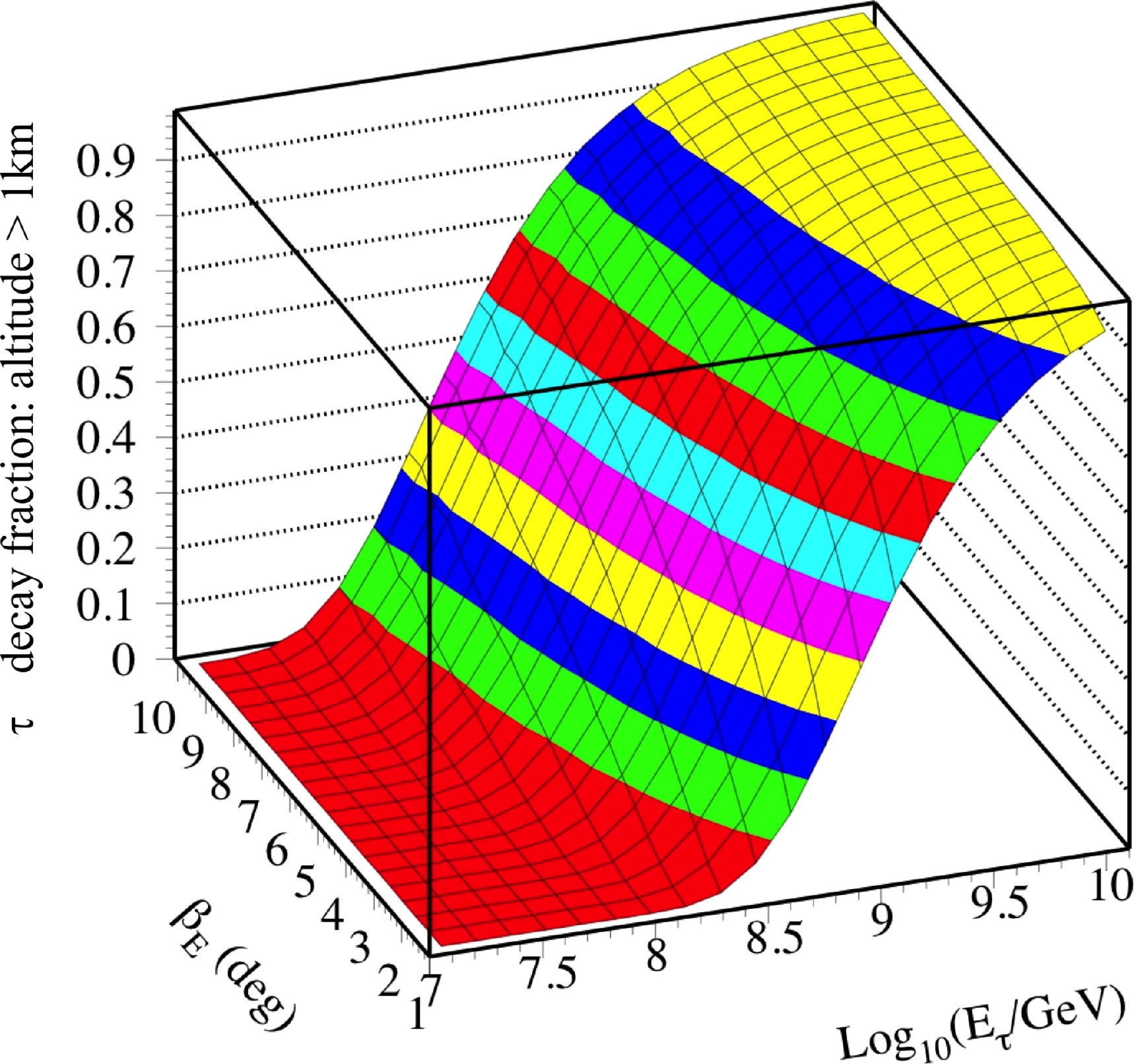}
\caption{The fraction of \taon decays above 1 km altitude as a function of \taon energy and Earth-emergence angle ($\beta_E$)}
\label{decayFrac}       
\end{figure}

Another effect imposing variability of the intensity EAS Cherenkov spectrum is the ability of \taons with energies above 100 PeV to decay and initiate EAS that develop above the aerosol layer.  Fig.~\ref{EASvsAlt} shows the results of a study model where 100 PeV EASs were started at different altitudes for $\beta_E=1^\circ$. The effects on the intensity and Cherenkov spectrum are more pronounced than that due to the variability for different $\beta_E$.  The span of 5 km to 10 km in EAS starting altitude shows nearly a 100-fold increase in the peak of the Cherenkov spectrum while starting the EAS at higher altitudes pushes the peak of the spectrum to lower wavelengths.  This variability is again dominated by the effects due to aerosol attenuation, which become more pronounced for smaller $\beta_E$. Fig.~\ref{decayFrac} shows the fraction of \taon decays that occur above 1 km, the nominal aerosol scale height, as a function of $\beta_E$ and \taon energy.   Note that to make a comparison to the spectrum in Fig.~\ref{cprofile} for $\beta_E=7^\circ$, one needs to take into account that the energy in an EAS is less than that of the decaying \taon.  
Using Pythia \cite{sjo15} to generate left-handed polarized \taon decays, one sees that using $E_{\rm EAS} \approx 0.5 E_\tau$ is a good average approximation \cite{2019PhRvD.100f3010R}. 

\section{Atmospheric NightGlow Background Effects on the Cherenkov Signal Threshold}

\begin{figure}[h]
\centering
\includegraphics[width=0.99\columnwidth]{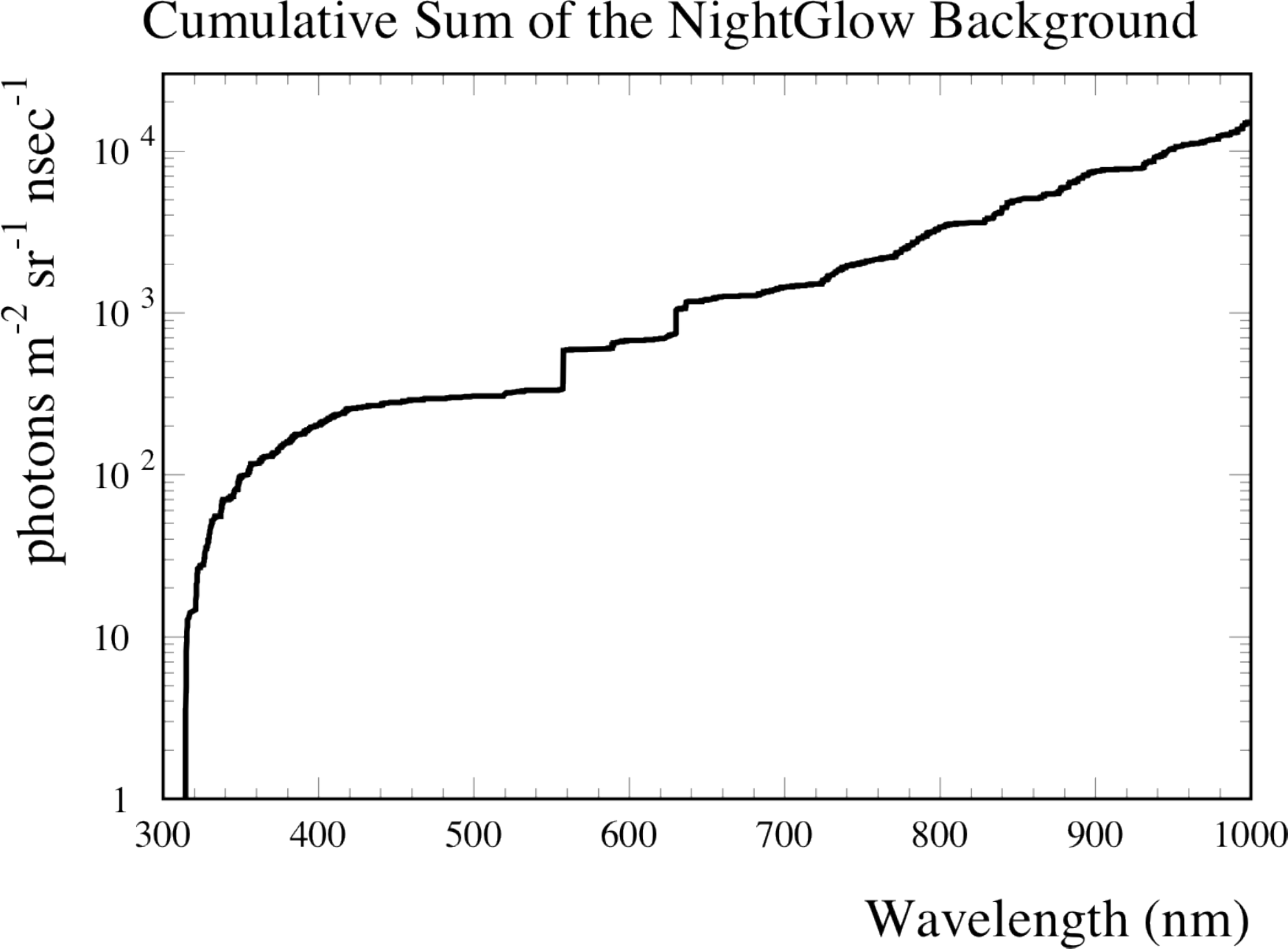}
\caption{The cumulative sum of the NightGlow Background (NGB) using the measurements of Hanuschik \cite{2003A&A...407.1157H,2006JGRA..11112307C}.}
\label{NGB}       
\end{figure}

The detection of both the air fluorescence and optical Cherenkov signals is significantly affected by the atmospheric nightglow background (NGB),  since measurements of the optical signals from EASs are performed during dark, nearly moonless nights.  Fig.~\ref{NGB} shows the cumulative sum of the NGB as a function of wavelength based on the measurements of Hanuschik \cite{2003A&A...407.1157H,2006JGRA..11112307C}. As is discussed in the next section, the EAS air fluorescence signal is 
dominated by lines below 500 nm.  The use of a UV filter matched to the wavelength response of a PMT allows the NGB background to be constrained to relatively modest values, $\sim 300$ photons m$^{-2}$ sr$^{-1}$ nsec$^{-1}$ given by the plateau around 500 nm in Fig.~\ref{NGB}. However, the variability of the optical Cherenkov spectrum generated by upward-moving \taon induced EASs and the desire to use the wide bandpass of SiPMs to optimize the Cherenkov detection translates into dealing with a NGB $\sim$15,000 photons m$^{-2}$ sr$^{-1}$ nsec$^{-1}$ out to 1000 nm.  The POEMMA focal plane count rate and probability of a false positive neutrino Cherenkov detection is determined by using an effective PDE by weighting the PDE versus wavelength response using the NGB spectrum, then calculating the cumulative Poisson probabilities as a function of the photo-electron (PE) threshold. Using the Hanuschik NGB spectrum from 314 -- 1000 nm and the measured PDE response of a Ham1520-6050CN SiPM an effective PDE of 9.7\% is calculated. Using this with the Hanuschik NGB rate of 15,000 photons m$^{-2}$ sr$^{-1}$ nsec$^{-1}$, a 10 ns integration time, and 2.5 m$^2$ effective area for POEMMA for the PCC, a threshold of 10 PEs yields a focal plane count rate $<1$ kHz for POEMMA. In the condition that the POEMMA satellites are separated by $\sim$25 km to co-measure the Cherenkov light pool for each event, using a 20 ns time coincidence yields a false positive `neutrino' detection rate of $<1$ event/year.  For the configuration where the satellites are separated by 300 km and individually view the Cherenkov signal from upward-moving \taon EASs, a PE threshold of 20 is needed to keep the false positive rate $< 1$ event/year.  Given the relatively high PE threshold values imposed by the NGB and the short integration time optimized for the Cherenkov signal, the effects of the dark count rate in SiPMs are mitigated. The POEMMA PE threshold calculations for the NGB also assume that effects of SiPM optical cross talk for the Cherenkov measurements are small compared to that of the NGB.

\section{SiPM use for EAS Air Fluorescence Measurement}

POEMMA uses stereo measurements of the isotropic air fluorescence generated by the EASs from UHECRs above 20 EeV to measure the evolution of the EAS longitudinal development with 1 $\mu$s time sampling. As previously discussed, the baseline design of POEMMA's PFC employs an MAPMT-based system with a BG3 UV filter to optimize the air fluorescence signal versus the NGB in the near UV (see the thicker magenta curve in Fig.~\ref{PDEcomp}).  Here the potential of using SiPMs to perform the air fluorescence EAS measurements is assessed, based on determining the wavelength response of MAPMT's and SiPMs to the air fluorescence spectrum.  The wavelength band of air fluorescence extends from below 200 nm to over 1000 nm \cite{Davidson}. However, the majority of the signal is in the wavelength band from 310 nm to 430 nm \cite{Davidson,Bunner1966}, which allows the use of UV filters to limit the effects of the NGB on the measurement.  A value of $\sim$ 500 photons m$^{-2}$ ns$^{-1}$ $sr^{-1}$ for the NGB in the 300 -- 400 nm wavelength band viewed from low-Earth orbit is based on measurements from balloon altitudes by NIGHTGLOW \cite{2005APh....22..439B}.  This is modestly higher than that obtained from the Hanuschik measurements, and the NIGHTGLOW NGB value is used to conservatively assess the effects on the threshold for UHECR EAS air fluorescence detection.

\begin{figure}[h]
\centering
\includegraphics[width=0.99\columnwidth]{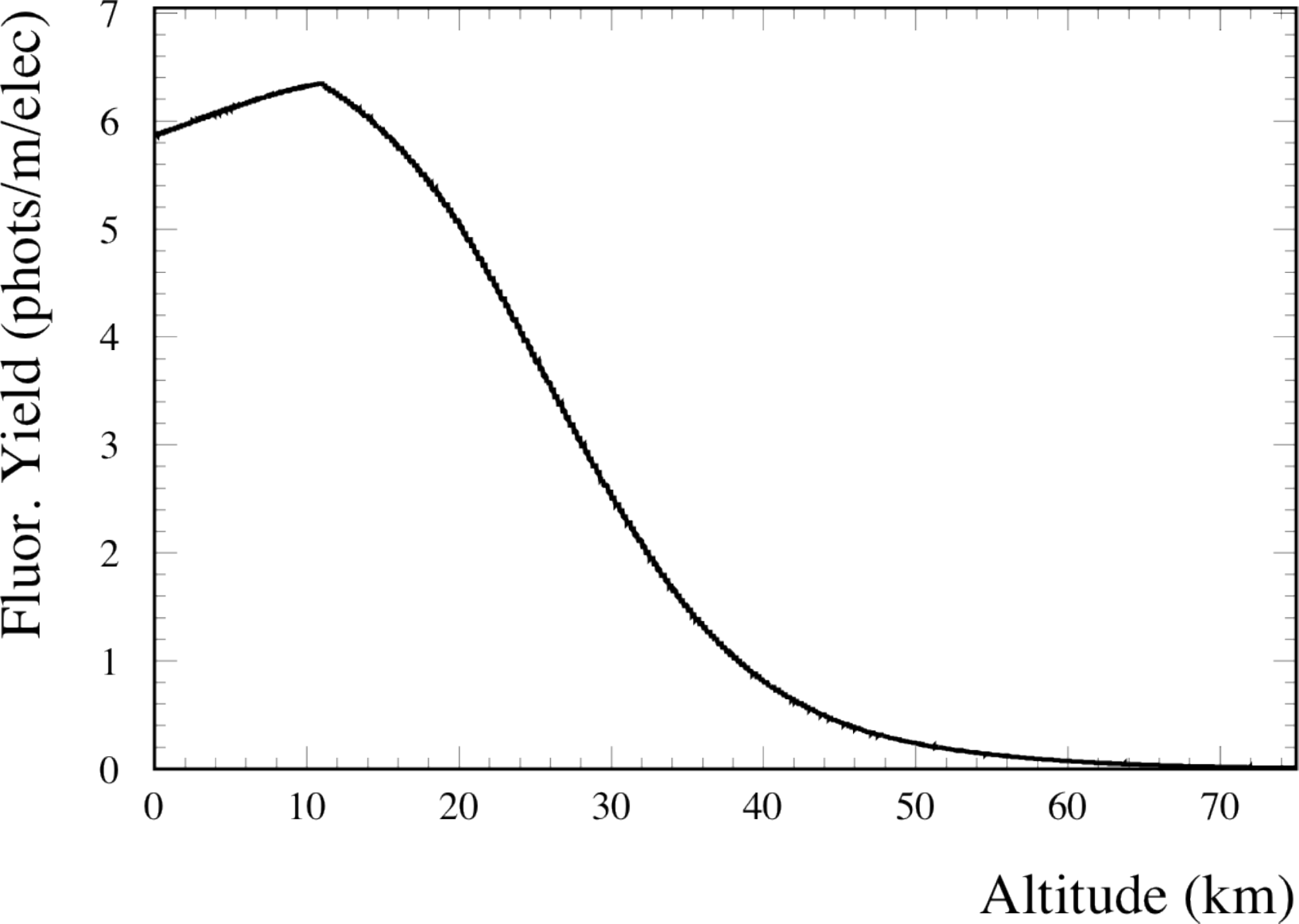}
\caption{The air fluorescence yield as a function of altitude.}
\label{FYalt}       
\end{figure}

Fig. \ref{FYalt} shows the air fluorescence yield as a function of altitude based on Kakimoto et al. \cite{1996NIMPA.372..527K} with the yield increased by 25\% to reflect the current average reported by PDG \cite{PDG}. For 50 EeV UHECRs observed by POEMMA in stereo mode, the mean altitude where the EAS maximum (\xmax) occurs is around 6 km.  The wavelength dependence of the fluorescence emission as measured by Bunner \cite{Bunner1966} is used, which has good agreement in the relative line strengths with more recent measurements using electron cascades in accelerator beams \cite{2008APh....29...77A}. Tab.~\ref{AirFluorComp} details the air fluorescence yields reported by Bunner around the principle lines both at EAS generation and at the POEMMA instruments after being attenuated through the atmosphere.  An atmospheric column depth of 550 g/cm$^2$ and ozone column 
depth of 330 milli-atm-cm was used for the attenuation calculations based on the viewed path to \xmax obtained from POEMMA stereo simulations of 50 EeV UHECRs. Note that since \xmax is around 6 km altitude for 50 EeV UHECRs, aerosol attenuation is negligible above this altitude for a nominal atmosphere. Note also for this comparison, the relative fluorescence line intensities are important, and the overall yield presented in the table is $\sim$25\% lower than the current PDG average value.

\begin{table}
\centering
\caption{Air Fluorescence Relative Yields per EAS charged particle.}
\label{AirFluorComp}       
{\footnotesize
\begin{tabular}{cccccc}
\hline

Peak $\lambda$  &   EAS Yield  & Atten Yield  & PMT  & FBK   & Ham14  \\ 
 & &   &  w/BG3 & w/FF01 & w/FF01 \\ 
 (nm) & (phots/m) &  (phots/m) &  (PEs/m) & (PEs/m) & (PEs/m) \\ \hline
315.9 & 0.43 & 0.24 & 0.09 & 0.04 & 0.02 \\
337.1 & 1.25 & 0.98 & 0.41 & 0.34 & 0.19 \\
357.7 & 1.17 & 0.95 & 0.40 & 0.34 & 0.23 \\
380.5 & 0.60 & 0.49 & 0.20 & 0.18 & 0.14 \\
391.4 & 0.65 & 0.54 & 0.22 & 0.29 & 0.25 \\
420.1 & 0.03 & 0.03 & 0.01 & 0.01 & 0.01 \\
427.0 & 0.17 & 0.14 & 0.04 & 0.06 & 0.06 \\
{\bf Total} & {\bf 4.29} & {\bf 3.36} & {\bf 1.38} & {\bf 1.26} & {\bf 0.90} \\ \hline
\end{tabular}
}
\end{table}

Also shown in Tab. ~\ref{AirFluorComp} are the photo-electron (PE) signals using the MAPMT QE and SiPM PDE responses shown in Fig.~\ref{PDEcomp} assuming a Schott BG3 filter is used with the MAPMT and a Semrock Brightline FF01-375/110-25 filter\footnote{Thanks to Pavel Klimov for calling attention to this UV filter.} \cite{Semrock} is used with the SiPMs. The Semrock filter has a 110 nm bandpass around 375 nm with virtually no transmission below or above the bandpass, as opposed to the BG3 filter.  
However, the Semrock filter is an interference filter with a half-cone angle of $7^\circ$ which is not be desirable for an optical system with a much larger FoV.\footnote{Thanks to Pat Reardon for this discussion point.} Nevertheless, the performance of the Semrock filter is evaluated as an example filter with minimal bandpass above 500 nm to reject the majority of the NGB.
Fig.~\ref{PDEcomp} shows the BG3 bandpass below 600 nm highlighted with a thicker curve than the transmission above this wavelength. As the plot indicated, this filter recovers to nearly 100\% transmission above 700 nm with significant transmission to and above 1000 nm. Thus, given the substantial intensity of the NGB above 700 nm, a BG3 filter would not be appropriate to use with a wide-bandpass photo-detector like a SiPM. Compared to the MAPMT w/BG3 result, the FBK SiPM w/FF01 records 92\% of the signal while the Hamamatsu Series 14 SiPM w/FF01 registers 65\% of the MAPMT signal. Not shown in the table are the results for a Hamamatsu Series 13 SiPM \cite{Ham13} w/FF01, which has 50\% of the MAPMT w/BG3 response. This comparison between the FBK SiPM and MAPMT responses indicates that this SiPM offers nearly identical air fluorescence response based on its measured PDE. While issues such as SiPM darkcount rate and optical crosstalk need to be considered, it is feasible to consider a SiPM-based focal plane for a space-based UHECR air fluorescence camera, which would yield significant mass reduction compared to a MAMPT-based system along with the elimination of the need for high voltage.

\section{Summary}

In this paper, the use of SiPMs to detect the Cherenkov light from cosmic neutrinos is detailed in the context of POEMMA. The broad wavelength range of SiPMs is well matched to the inherent variability of the Cherenkov spectrum generated by upward-moving EASs from Earth-emergent \taon decays generated by tau neutrino interactions in the Earth.  While the significant PDE for SiPMs at longer wavelength aids the sensitivity to lower energy cosmic neutrinos, the response at shorter wavelengths is equally important considering that the peak of the \taon EAS Cherenkov emission moves to shorter wavelengths for both higher energy 
neutrinos and larger Earth-emergence angles since more or the EAS development occurs above the aerosol layer. In particular, the increase in effective PDE for larger Earth-emergence angles is important for neutrino ToO observations when the POEMMA satellites are following the transient sources. A comparison of the air fluorescence response of SiPMs with significant PDE in the near UV demonstrates that, at least from a photo-detection standpoint, currently available SiPMs have nearly identical response as MAPMTs, but each have specific UV filter requirements. Thus the potential exists for space-based SiPM-based instruments tuned to the EAS air fluorescence signal with less massive focal planes and without the need for high voltage. 

\section{Acknowledement}

I would like to thank the POEMMA collaboration and in particular Nepomuk Otte for providing the SiPM measurements and Simon Mackovjak for assistance with the NGB evaluation.  The work presented in this paper was supported by NASA grants NNX17AJ82 and 80NSSC19K0626.

\vspace{3 mm}
\noindent $\dagger$ {\bf POEMMA Collaboration} A. V. Olinto$^1$, J. H. Adams$^2$, R. Aloisio$^3$, L. A.
Anchordoqui$^4$, M. Bagheri$^5$, D. Barghini$^6$, M. Battisti$^6$, D. R. Bergman$^7$, M. E. Bertaina$^6$,  P. Bertone$^8$, F. Bisconti$^9$, M. Bustamante$^{10}$, M. Casolino$^{11,12}$, M. J.
Christl$^8$, A. L. Cummings$^3$, I. De Mitri$^3$, R. Diesing$^1$, R.Engel$^{13}$, J. Eser$^{14}$, K. Fang$^{15}$, G. Filippatos$^{14}$, F. Fenu$^6$, E. Gazda$^5$, C. Guepin$^{16}$, E. A. Hays$^{17}$, E. G. Judd$^{18}$, P. Klimov$^{19}$, J. F. Krizmanic$^{17,20}$, V. Kungel$^{14}$, E.
Kuznetsov$^2$, S. Mackovjak$^{21}$, L. Marcelli$^{12}$, J. McEnery$^{17}$,K.D. Merenda$^{14}$,  S. S. Meyer$^1$, J. W. Mitchell$^{17}$, H. Miyamoto$^6$, J. M. Nachtman$^{22}$, A. Neronov$^{23}$, F. Oikonomou$^{24}$, Y. Onel$^{22}$, A. N. Otte$^5$,
E. Parizot$^{25}$, T. Paul$^4$,
J. S. Perkins$^{17}$, P. Picozza$^{12,26}$, L.W. Piotrowski$^{11}$, Z. Plebaniak$^{28}$, G. Pr\'ev\^ot$^{25}$, P. Reardon$^2$, M. H. Reno$^{22}$, M.Ricci$^{27}$,O. Romero Matamala$^5$, F. Sarazin$^{14}$, K. Shinozaki$^{28}$, J. F. Soriano$^4$, F. Stecker$^{17}$, Y. Takizawa$^{11}$, R. Ulrich$^{13}$, M. Unger$^{13}$, T. Venters$^{17}$, L.
Wiencke$^{14}$, D. Winn$^{22}$, R. M. Young$^8$, M. Zotov$^{19}$

 {\it $^1$The University of Chicago, Chicago, IL, USA;
$^2$University of Alabama, Huntsville, AL, USA;
$^3$Gran Sasso Science Institute, L'Aquila, Italy;
$^4$CUNY, Lehman College, NY, USA; 
$^5$Georgia Institute of Technology, Atlanta, GA, USA; 
$^6$Universit\`{a} di Torino, Torino, Italy; 
$^7$University of Utah, Salt Lake City, Utah, USA; 
$^8$NASA Marshall Space Flight Center, Huntsville, AL, USA; 
$^9$Istituto Nazionale di Fisica Nucleare, Section of Turin, Turin, Italy;
$^{10}$Niels Bohr Institute, University of Copenhagen, DK-2100 Copenhagen, Denmark;
$^{11}$RIKEN, Wako, Japan;
$^{12}$Istituto Nazionale di Fisica Nucleare, Section of Roma Tor Vergata, Italy;
$^{13}$Karlsruhe Institute of Technology, Karlsruhe, Germany;
$^{14}$Colorado School of Mines, Golden, CO, USA;
$^{15}$Kavli Institute for Particle Astrophysics and Cosmology, Stanford University, Stanford, CA, USA
$^{16}$Department of Astronomy, University of Maryland, College Park, MD, USA;
$^{17}$NASA Goddard Space Flight Center, Greenbelt, MD, USA;
$^{18}$Space Sciences Laboratory, University of California, Berkeley, CA, USA;
$^{19}$Skobeltsyn Institute of Nuclear Physics, Lomonosov Moscow State University, Moscow, Russia;
$^{20}$Center for Space Science \& Technology, University of Maryland, Baltimore County, Baltimore, MD, USA;
$^{21}$Institute of Experimental Physics, Slovak Academy of Sciences, Kosice, Slovakia;
$^{22}$University of Iowa, Iowa City, IA, USA;
$^{23}$University of Geneva, Geneva, Switzerland; 
$^{24}$Institutt for fysikk, NTNU, Trondheim, Norway;
$^{25}$Universit\'e de Paris, CNRS, Astroparticule et Cosmologie, F-75013 Paris, France;
$^{26}$Universita di Roma Tor Vergata, Italy;
$^{27}$Istituto Nazionale di Fisica Nucleare - Laboratori Nazionali di Frascati, Frascati, Italy;
$^{28}$National Centre for Nuclear Research, {\L}\'{o}d\'{z, Poland}

\section*{References}

\bibliography{Baribibfile}

\end{document}